\documentclass[]{aa}
\usepackage{graphicx}
\usepackage[varg]{txfonts}
\usepackage{hyperref}
\usepackage[version=3]{mhchem}

\begin{document} 

\title{Clouds in the atmospheres of extrasolar planets}
\subtitle{V. The impact of \ce{CO2} ice clouds on the outer boundary of the habitable zone}

\titlerunning{Clouds in the atmospheres of extrasolar planets. V.}

\author{D. Kitzmann\inst{1}}
\institute{Physikalisches Institut \& Center for Space and Habitability, University of Bern,
           Sidlerstr. 5, 3012 Bern, Switzerland\\
           \email{daniel.kitzmann@csh.unibe.ch}}

 \date{Received 8 November 2016 / Accepted 25 January 2017}
  
 \abstract{Clouds have a strong impact on the climate of planetary atmospheres. The potential scattering greenhouse effect of CO$_2$ ice clouds in the atmospheres of terrestrial extrasolar planets is of particular interest because it might influence the position and thus the extension of the outer boundary of the classic habitable zone around main sequence stars. Here, the impact of CO$_2$ ice clouds on the surface temperatures of terrestrial planets with CO$_2$ dominated atmospheres, orbiting different types of stars is studied. 
 Additionally, their corresponding effect on the position of the outer habitable zone boundary is evaluated. For this study, a radiative-convective atmospheric model is used the calculate the surface temperatures influenced by CO$_2$ ice particles. The clouds are included using a parametrised cloud model. The atmospheric model includes a general discrete ordinate radiative transfer that can describe the anisotropic scattering by the cloud particles accurately. 

 A net scattering greenhouse effect caused by CO$_2$ clouds is only obtained in a rather limited parameter range which also strongly depends on the stellar effective temperature. For cool M-stars, \ce{CO2} clouds only provide about 6 K of additional greenhouse heating in the best case scenario. On the other hand, the surface temperature for a planet around an F-type star can be increased by 30 K if carbon dioxide clouds are present.
 Accordingly, the extension of the habitable zone due to clouds is quite small for late-type stars. Higher stellar effective temperatures, on the other hand, can lead to outer HZ boundaries about 0.5 au farther out than the corresponding clear-sky values.}


   \maketitle

\section{Introduction}

Clouds can have an important impact on the climate of terrestrial planets by either trapping the infrared radiation in the lower atmosphere (greenhouse effect) or by scattering incident stellar 
radiation back to space (albedo effect). The position and extension of the habitable zone (HZ) around different types of stars thus depends on the presence of clouds \citep{Marley2013cctp.book..367M}. 
Especially, the outer HZ boundary might be influenced by the formation of CO$_2$ ice clouds and their corresponding climatic impact. If a planet is located farther away from its host star, its 
atmospheric and surface temperatures become cooler due to the decrease in stellar insolation. To sustain liquid water on the surface, a thick atmosphere composed of a greenhouse gas, such as \ce{CO2}, is required. If the terrestrial planet is still geologically active, CO$_2$ can accumulate in the atmosphere by volcanic outgassing \citep[see][for an overview]{Pierrehumbert2010ppc..book.....P}. With decreasing atmospheric temperatures, carbon dioxide will condense at some point to form clouds composed of CO$_2$ ice crystals.

In contrast to other types of condensates important for habitable, terrestrial planets -- such as liquid \ce{H2O} or water ice -- dry ice is more or less transparent in the infrared except within a few strong absorption bands \citep{Hansen1997JGR,Hansen2005JGRE} (see the corresponding refractive index in Fig. 1 in \citet{Kitzmann2013A&A...557A...6K}).
Thus, as argued by \citet{Kasting1993} or \citet{Forget1997}, a classical greenhouse effect by absorption and re-emission of thermal radiation is unlikely to occur for \ce{CO2} ice clouds.
However, as pointed out by \citet{Forget1997} and \citet{Pierrehumbert1998JAtS}, CO$_2$ ice particles can efficiently scatter thermal radiation back to the planetary surface thereby creating a scattering greenhouse effect. Depending on the cloud properties a scattering greenhouse effect can outweigh the cloud's albedo effect and can, in principle, increase the surface temperature above the freezing point of water \citep[see also][for studies of CO$_2$ clouds in the early Martian atmosphere]{Mischna2000Icar,Colaprete2003JGRE}.

\begin{figure}
  \includegraphics[scale=0.57]{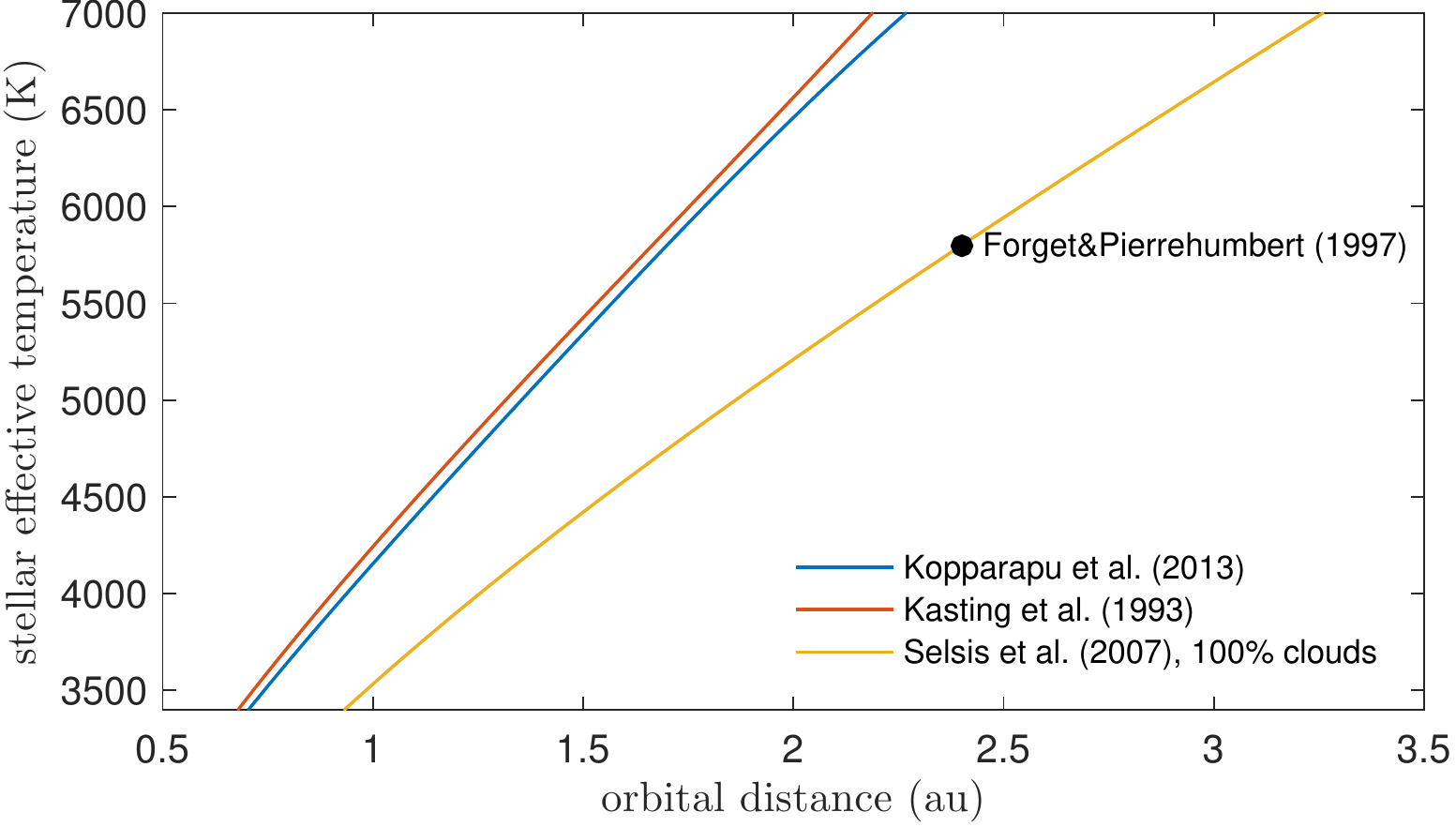}
  \caption{Estimates for the outer boundary of the classical habitable zone from previous studies. The red and blue curves show the cloud-free boundaries from \citep{Kasting1993} and 
\citet{Kopparapu2013ApJ...765..131K}, respectively. The results for an atmosphere with 100\% \ce{CO2} clouds from \citet{Selsis2007} is shown in yellow. This estimate is based on a calculation by 
\citet{Forget1997}, which is additionally marked in the figure.}
  \label{fig:hz_old}
\end{figure}

Most atmospheric modelling studies on the climatic effects of CO$_2$ clouds so far have been limited to the early Martian atmosphere. For a fully cloud-covered early Mars with a thick CO$_2$ dominated 
atmosphere and CO$_2$ clouds composed of spherical CO$_2$ ice particles, \citet{Forget1997} determined that in contrast to the cloud-free distance of 1.67 au by \citet{Kasting1993}, the outer 
boundary of the HZ should be located at 2.4 au. 
This value has been further used by \citet{Selsis2007} to extrapolate the effects of CO$_2$ clouds on the outer HZ boundary towards other main sequence central 
stars, assuming that the radiative effects of CO$_2$ clouds are not a function of the incident stellar radiation or the properties of the planet and its atmosphere. 
The cloud-free outer HZ has recently been slightly revised by \citet{Kopparapu2013ApJ...765..131K} who used an updated version of the \citet{Kasting1993} model. 
The corresponding results of these earlier studies on the outer boundary of the classical habitable zone are summarised in Fig. \ref{fig:hz_old}.

For the super-Earth Gliese 581 d orbiting the M3V dwarf Gliese 581, a simplified description of \ce{CO2} ice particle formation was included in the one-dimensional (1D) and three-dimensional (3D) atmospheric models by \citet{Wordsworth2010A&A...522A..22W,Wordsworth2011ApJ}. These studies stated that the \ce{CO2} clouds contribute to the greenhouse effect by increasing the planet's surface temperature. However, they also employed a simplified treatment of the radiative transfer by using two-stream methods.

The first extensive study on the radiative effect of CO$_2$ clouds in atmospheres of terrestrial planets around main-sequence dwarf stars was done by \citet{Kitzmann2013A&A...557A...6K}. In 
that radiative transfer study, the properties of the CO$_2$ ice particles (particle sizes, optical depths) were varied over a large parameter range to calculate their radiative effects using a 
high-order discrete ordinate radiative transfer method. The results of this study suggest, that the simplified two-stream radiative transfer methods employed in previous model studies strongly 
overestimated the scattering greenhouse effect of CO$_2$ clouds, concluding that more accurate radiative transfer schemes are absolutely required for atmospheric models to predict the correct climatic 
impact of CO$_2$ ice particle clouds.

Following this numerical radiative transfer study, \citet{Kitzmann2016ApJ...817L..18K} reinvestigated the impact of \ce{CO2} ice clouds in the atmosphere of early Mars by using a radiative-convective 
atmospheric model with an accurate radiative transfer. While the results in \citet{Kitzmann2016ApJ...817L..18K} suggest, that carbon dioxide clouds still yield a net greenhouse effect under certain 
conditions, the impact on the surface temperature is much less pronounced than found in the previous studies on this topic. This reduced heating effect will also affect the previous 
estimates on the extension of the HZ due to the presence of \ce{CO2} clouds.

In this work, I study the climatic effect of CO$_2$ ice clouds in CO$_2$-dominated atmospheres of terrestrial planets around different types of main-sequence dwarf stars by using a 1D 
radiative-convective atmospheric model with an accurate multi-stream radiative transfer. In particular, the impact of carbon dioxide clouds on the position of the outer boundary of the classical 
habitable zone is investigated. 
Section \ref{sec:model_description} gives an overview of the atmospheric model used in this study, as well as the CO$_2$ 
cloud description. 
The climatic impact of CO$_2$ clouds is discussed in Sect. \ref{sec:cloud_climatic_impact}, while their effect on the outer HZ boundary is studied in Sect. \ref{sec:co2_hz}.
Concluding remarks and a summary are given in Sect. \ref{sec:summary}.

\section{Model description}
\label{sec:model_description}

For the calculations in this publication, I use a 1D radiative-convective atmospheric model, previously used to study the climatic impact of \ce{CO2} clouds in the atmosphere of early Mars.
A detailed model description is presented in the following.

The model features a state-of-the-art radiative transfer treatment based on opacity sampling and a general discrete ordinate method, able to accurately treat 
anisotropic scattering. The model currently considers \ce{N2}, \ce{CO2}, and \ce{H2O} as atmospheric species. 

In the dry atmosphere, \ce{N2} and \ce{CO2} are considered to be well mixed, that is, they have a constant mixing ratio throughout the atmosphere. For water, the relative humidity profile of 
\citet{Manabe67} is used, with a fixed relative humidity at the surface of 77\%. \citet{Godolt2016A&A...592A..36G} showed that the results using this relative humidity profile are a good 
approximation in comparison to full 3D studies of cold atmospheres. 

The atmospheric model is stationary, that is, it doesn't contain an explicit time dependence. It does, however, use the usual approach of time stepping to calculate the atmospheric temperatures 
\citep[e.g.][]{Kasting1993}. The radiative equilibrium temperatures are obtained via
\begin{equation}
  \frac{\mathrm d}{\mathrm d t} T(z) = -\frac{g}{c_p(z)} \frac{\mathrm d F(z)}{\mathrm d p(z)} \ ,
  \label{eq:time_stepping}
\end{equation}
where $g$ is the gravitational acceleration, $c_p$ the heat capacity at constant pressure and $F$ the wavelength-integrated radiation flux.
It should be noted that the time $t$ in Eq. \eqref{eq:time_stepping} needs not to be treated as a real time variable. It is, rather, just an iteration parameter to drive the atmosphere into 
equilibrium.
For the present model, $t$ is not a global constant, but is allowed to change as a function of grid point.
This allows, for example, to use relatively large values of $t$ at grid points with very high thermal inertia (usually at high pressures), whereas it can be smaller at the top of the atmosphere to 
stabilise the convergence in this region.

In regions where the atmosphere is found to be convectively unstable, convective adjustment is performed. The convective lapse rate for regions with \ce{CO2} and \ce{H2O} condensation is modelled 
after \citet{Wordsworth2013ApJ...778..154W}.

Approximately one hundred grid points are used to discretise the vertical extension of the atmosphere. For the surface albedo, I use the mean Earth-like value of 0.13 \citep{Kitzmann2010A&A...511A..66K}.

\subsection{Radiative transfer}

A single radiative transfer scheme with a multi-stream discrete ordinate method is used for the entire wavelength range from 0.1 $\mu$m to 500 $\mu$m.
This method solves the transfer equation
\begin{equation}
\mu \frac{\mathrm{d} I_\lambda (\tau_\lambda, \mu)}{\mathrm{d}\tau_\lambda} = I_\lambda(\tau_\lambda, \mu) - S_{\lambda}(\tau_\lambda, \mu) 
\label{eq:rte}
\end{equation}
at several distinct values of the angular variable $\mu$ (streams) and afterwards computes the averaged radiation field quantities (radiation flux, mean intensity) by angular integration. 
The general source function $S_{\lambda}(\tau_\lambda)$ takes the incident radiation from the central star ($S_{\lambda,\mathrm{*}}$), the local 
thermal emission, described by the Planck function $B_\lambda$, and the contributions due to scattering into account. It is given by
\begin{equation}
\begin{split}
S_{\lambda}(\tau_\lambda) = & S_{\lambda,\mathrm{*}}(\tau_\lambda) + (1 - \omega_\lambda) B_\lambda\\  
&   + \frac{\omega_\lambda}{2} \int_{-1}^{+1} p_\lambda^0(\mu,\mu')I_\lambda(\mu') \mathrm{d}\mu' \ ,
\end{split}
\end{equation}
where $\omega_\lambda$ is the single scattering albedo and $p_\lambda^0(\mu,\mu')$ is the azimuthally-averaged scattering phase function.

Here, the scattering phase function is represented as a series of Legendre polynomials \citep{Chandrasekhar1960ratr}
\begin{equation}
  p_\lambda^0(\mu,\mu') = \sum_{n=0}^{\infty} (2 n + 1) P_n(\mu) P_n(\mu') \chi_{\lambda,n}
  \label{eq:phase_function}
\end{equation}
with the Legendre polynomials $P_n(\mu)$ and the phase function moments $\chi_{\lambda,n}$. 
The moments are defined by the integration of the full phase function $p_\lambda(\alpha)$ with respect to the scattering angle $\alpha$, weighted by the Legendre polynomials $P_n(\mu)$
\begin{equation}
  \chi_{\lambda,n} = \frac{1}{2} \int_{-1}^{+1} p_\lambda(\alpha) P_n (\cos \alpha) \mathrm d \cos \alpha \ .
\end{equation}
For most phase functions, this series is infinite but is, in practice, truncated at a certain
$n=N_\mathrm{max}$. 
As mentioned by \citet{Chandrasekhar1960ratr}, the value of $N_\mathrm{max}$ is equal to the number of streams used to solve the radiative transfer equation.

To treat the strong forward scattering peak of the phase function in case of large size parameters, the $\delta$-M scaling method from \citet{Wiscombe1977JAtS...34.1408W} is used. 
This method approximates the forward scattering peak by a $\delta$ distribution and removes it from the phase function series.
The $\delta$-M scaling, however, produces errors in the computed intensities which are therefore corrected by using the method presented in \citet{Nakajima1988JQSRT..40...51N}. 

For the numerical implementation of the discrete ordinate method, I use the C-DISORT radiative transfer code \citep{Hamre2013AIPC.1531..923H}.
Throughout this study, eight computational streams are used in all calculations.
Tests by doubling the number of streams showed no changes in the resulting atmospheric and surface temperatures or radiation fluxes.

\subsubsection{Opacity sampling}

For the description of wavelength-dependent transport coefficients, I adopt the opacity sampling method. This method has been introduced in the context of the cool atmospheres of late-type stars 
which are dominated by molecular absorption \citep{Sneden1976ApJ...204..281S}.
It has the advantage of operating in the normal wavelength/wavenumber space, which allows the transport coefficients of different species to be directly added \citep{Mihalas1978stat.book.....M}. 
The rationale for employing opacity sampling is based on the fact, that wavelength-averaged quantities, such as the total radiation flux or the mean intensity, converge well before all spectral lines are fully resolved.

In the opacity sampling approach, the equation of radiative transfer is solved at distinct wavelength points, at which its results are identical to a line-by-line radiative transfer method. There are a number of strategies on how to distribute these distinct wavelengths, such that, for example, the wavelength-integrated flux converges for a small number of points.

As mentioned in \citet{Kitzmann2016ApJ...817L..18K}, the distribution of wavelengths at which the equation of radiative transfer is solved, is treated separately in three different wavelength regions. In the infrared, the points are sampled along the Planck black body curves for different temperatures, adopted from the method published by \citet{Helling1998A&A}.
Here, the points are sampled for 30 temperatures between 100 K and 400 K, covering the entire range of atmospheric temperatures encountered in the atmospheric scenarios of this study.
Between 0.3 $\mu$m and 5 $\mu$m, 20,000 wavelength points are distributed logarithmic equidistantly in wavenumbers. For smaller wavelengths, approximately 100 points are used to treat the smooth Rayleigh slope.

\begin{figure}
  \includegraphics[scale=0.57]{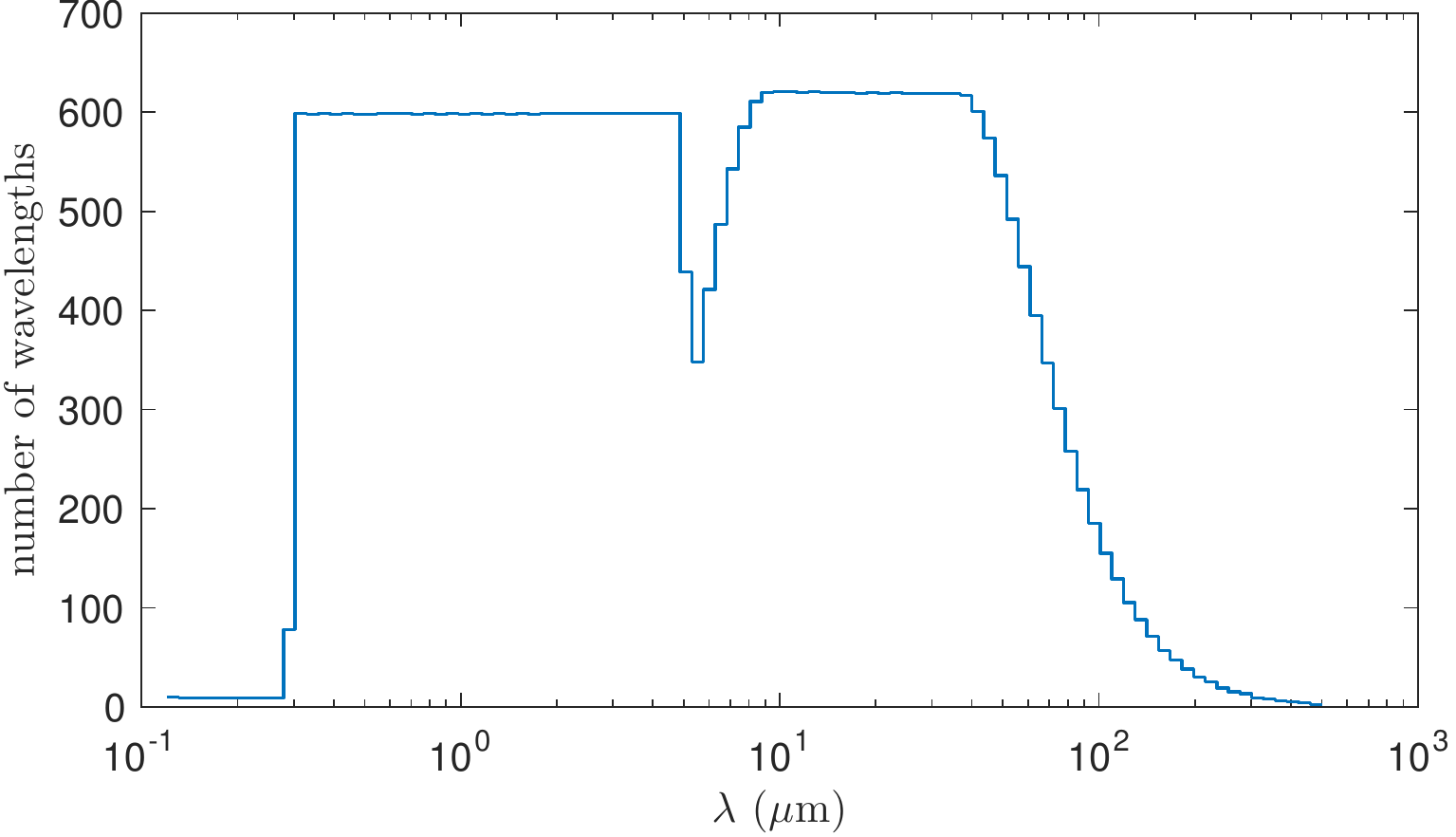}
  \caption{Histogram of the distribution of opacity sampling points as a function of wavelength. The histogram shows the number of wavelength points in 100 bins, equidistantly distributed in 
log space. The total number of sampling points is approximately 40,000.}
  \label{fig:resolution}
\end{figure}

Figure \ref{fig:resolution} shows a histogram of the distribution of opacity sampling points as a function of wavelength. 
The figure nicely illustrates the high sampling rate in the thermal infrared where most of the atmospheric infrared radiation is transported and the drop-off at the flanks of the high and low temperature Planck curves. 

In total approximately 40,000 distinct wavelengths are used in this study. Further increasing the spectral resolution has no impact on the atmospheric fluxes and temperature profiles.
In principle, the number of points could also be reduced significantly without any large impact on the temperatures. 
For example, reducing the wavelengths in the visible and near infrared by 10,000 yields changes in the surface temperatures of only 0.1 K. 
In fact, reasonable values for the surface temperatures can already be obtained by using only a couple of hundred wavelengths.

\subsubsection{Absorption coefficients}

The molecular absorption coefficients used in this study are calculated with the open source \textsc{Kspectrum} code (version 1.2.0). It should be noted that the currently available version of \textsc{Kspectrum} contains a bug in the calculation of the sub-Lorentzian line profiles for \ce{CO2}. A fixed version of the code has been forwarded to the code's author but has not been made publicly available so far.
Using the HITRAN 2012 database \citep{Rothman2013JQSRT.130....4R}, absorption cross-sections of \ce{CO2} and \ce{H2O} are obtained for pressures between $10^{-6}$ bar and $300$ bar and 
temperatures between 100 K and 640 K. The cross-sections are tabulated in equidistant wavenumber steps of 0.01 cm$^{-1}$.
The opacity sampling points are always chosen to coincide with one of the tabulated wavenumbers, thus avoiding spectral interpolation of the cross-sections.

In the case of \ce{CO2}, the sub-Lorentzian line profiles of \citet{Perrin1989JQSRT..42..311P} are employed. The line profiles are truncated at a distance of 500 cm$^{-1}$ from the line centre. The continuum contribution and dimer absorption between 1100 and 2000 cm$^{-1}$ is taken from 
\citet{Baranov2003JMoSp.218..260B} while the descriptions in \citet{Gruszka1997Icar..129..172G} are used to treat collision-induced absorption for wavenumbers smaller than 250 cm$^{-1}$. 
For the self and foreign continuum absorption of \ce{H2O}, the MT-CKD formulation \citep{Mlawer2012RSPTA.370.2520M} is used. Following the requirements of the MT-CKD model, the line wings of \ce{H2O} are truncated at distances of 25 cm$^{-1}$.

Recently, \citet{Ozak2016JGRE..121..965O} showed the importance of taking into account the line-coupling of the \ce{CO2} absorption lines in the infrared for the climate of the early Mars. The use of purely Lorentzian far wing line shapes can lead to an overestimation of the \ce{CO2} greenhouse effect in the infrared. The effect of line mixing is partly included in the model by using the sub-Lorentzian line profiles, though it may still overestimate the greenhouse warming by the \ce{CO2} gas under certain conditions. Thus, the HZ boundaries presented here could be an upper limit.

\subsubsection{Molecular Rayleigh scattering}

Molecular Rayleigh scattering \citep{doi:10.1080/14786449908621276} is included for \ce{CO2}, \ce{H2O}, and \ce{N2}.
The corresponding cross-section is computed via
\begin{equation}
  \sigma_{\mathrm{rayleigh},\nu} = \frac{24 \pi^3 \nu^4}{n_\mathrm{ref}^2} \cdot \left(\frac{n(\nu)^2 - 1}{n(\nu)^2 + 2}\right)^2 \cdot K(\nu) \ ,
\end{equation}
where $\nu$ is the wavenumber, $n$ the refractive index, $n_\mathrm{ref}$ a reference particle number density, and $K$ the King factor. The King 
factor describes a correction to account for anisotropic molecules. It can also be written as a 
function of the depolarization factor $D$
\begin{equation}
  K(\nu) = \frac{6 + 3 D(\nu)}{6 - 7 D(\nu)} \ .
\end{equation}
For water, the refractive index from \citet{wagner2008international} and the depolarisation factor of $3\cdot 10^{-4}$ from \citet{Murphy1977} are adopted.
The refractive indices and King factors for \ce{CO2} and \ce{N2} are taken from \citet{Sneep2005JQSRT..92..293S}. Note that Eq. (13) in \citet{Sneep2005JQSRT..92..293S} for the refractive 
index of \ce{CO2} contains a typographical error. The nominator of the last term should read $0.1218145\cdot 10^{-6}$ instead of the $0.1218145\cdot 10^{-4}$ factor stated in their Eq. (13).

\subsection{Cloud description}

The cloud description is identical to the one also used by \citet{Forget1997}, \citet{Kitzmann2013A&A...557A...6K}, or \citet{Kitzmann2016ApJ...817L..18K}.
The size distribution of the \ce{CO2} ice particles is described by a modified gamma distribution
\begin{equation}
  f(a) = \frac{\left( a_\mathrm{eff} \sigma \right)^{2-1/\sigma}}{\Gamma\left(\frac{1-2\sigma}{\sigma}\right)} a^{\left(\frac{1}{\sigma}-3\right)} \mathrm 
e^{-\frac{a}{a_\mathrm{eff}\sigma}} \,
  \label{eq:gamma_distribution}
\end{equation}
where $a_\mathrm{eff}$ is the effective particle radius and $\sigma$ the effective variance, for which a value of 0.1 is used \citep{Forget1997}. 

The optical properties of the ice particles are obtained via Mie theory, thereby assuming spherical particles \citep{Mie1908AnP}. The refractive index for dry ice is taken from 
\citet{Hansen1997JGR,Hansen2005JGRE}. 
A plot with the optical properties for some selected values of the effective radius and an optical depth 
of one can be found in \citet{Kitzmann2016ApJ...817L..18K}. Throughout this study, the cloud optical depth $\tau$ refers to the particular wavelength of $\lambda = 0.1 \ 
\mathrm{\mu m}$.

The scattering phase function of the \ce{CO2} ice cloud particles is approximated by the analytical Henyey-Greenstein function \citep{Henyey1941ApJ....93...70H}
\begin{equation}
p_{\mathrm{HG},\lambda}(g_\lambda,\alpha) = \frac{1-g_\lambda^2}{\left(1 + g_\lambda^2 - 2g_\lambda\cos\alpha \right)^{3/2} } \ ,
\end{equation}
where $g_\lambda$ is the asymmetry parameter obtained from Mie theory.

This function lacks the complicated structure and detailed features of the full Mie phase function, though preserves its average quantities, such as, most notably, the 
asymmetry parameter $g_\lambda$, that is
\begin{equation}
g_{\mathrm{HG},\lambda} = \frac{1}{2}  \int_{-1}^{+1} p_{\mathrm{HG},\lambda}(g_\lambda,\alpha) \cos\alpha \, \mathrm d \cos\alpha = g_\lambda \ .
\end{equation}
In case of the Henyey-Greenstein function, the phase function moments required for the Legendre series \eqref{eq:phase_function} have the simple form
\begin{equation}
   \chi_{\lambda,n} = g_\lambda^n \ .
\end{equation}

\begin{figure}
  \centering
  \includegraphics[width=0.35\textwidth]{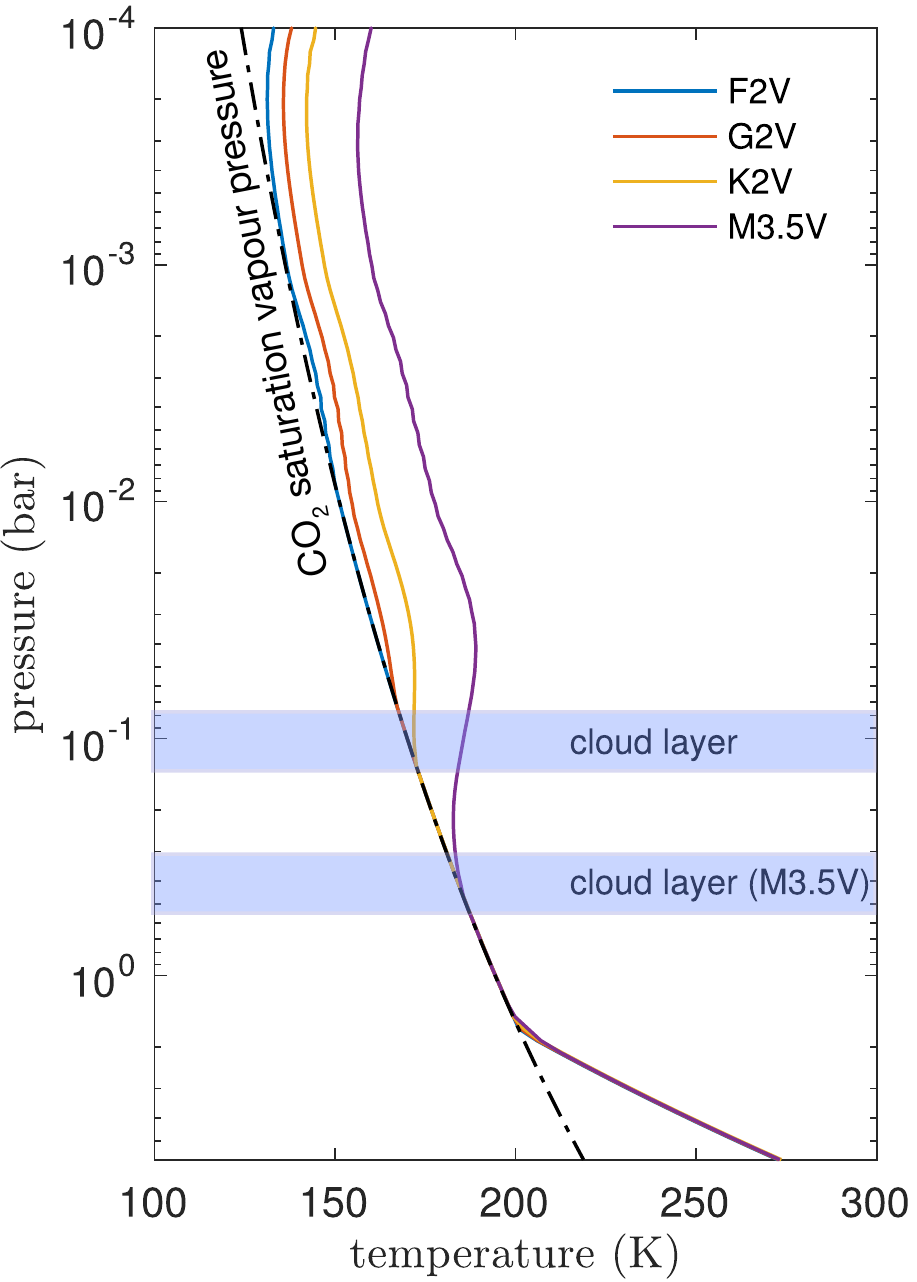}
  \caption{Temperature-pressure profiles for planets with cloud-free atmospheres around four different main-sequence stars. The corresponding atmospheres are composed of six bar of \ce{CO2} gas. The 
black, dashed-dotted line shows the saturation vapour pressure curve of \ce{CO2}. The areas coloured in blue indicate the positions where \ce{CO2} cloud layers are placed in the subsequent 
calculation.}
  \label{fig:clear_sky_profiles}
\end{figure}

\subsection{Stellar spectra}

Stellar spectra for four main-sequence stars with different stellar effective temperatures $T_\mathrm{eff}$ are used in this study.
This includes $\sigma$ Boo (F2V, $T_\mathrm{eff} = 6722$ K), the Sun (G2V, $T_\mathrm{eff} = 5777$ K), the young K-dwarf $\epsilon$ Eri (K2V, $T_\mathrm{eff} = 5072$ K), and AD Leo (M3.5eV, 
$T_\mathrm{eff} = 3400$ K).
The spectra are a composite of stellar atmosphere models and measurements. Details on the spectra can be found in \citet{Kitzmann2010A&A...511A..66K}.

\begin{figure*}
  \centering
  \includegraphics[scale=0.47]{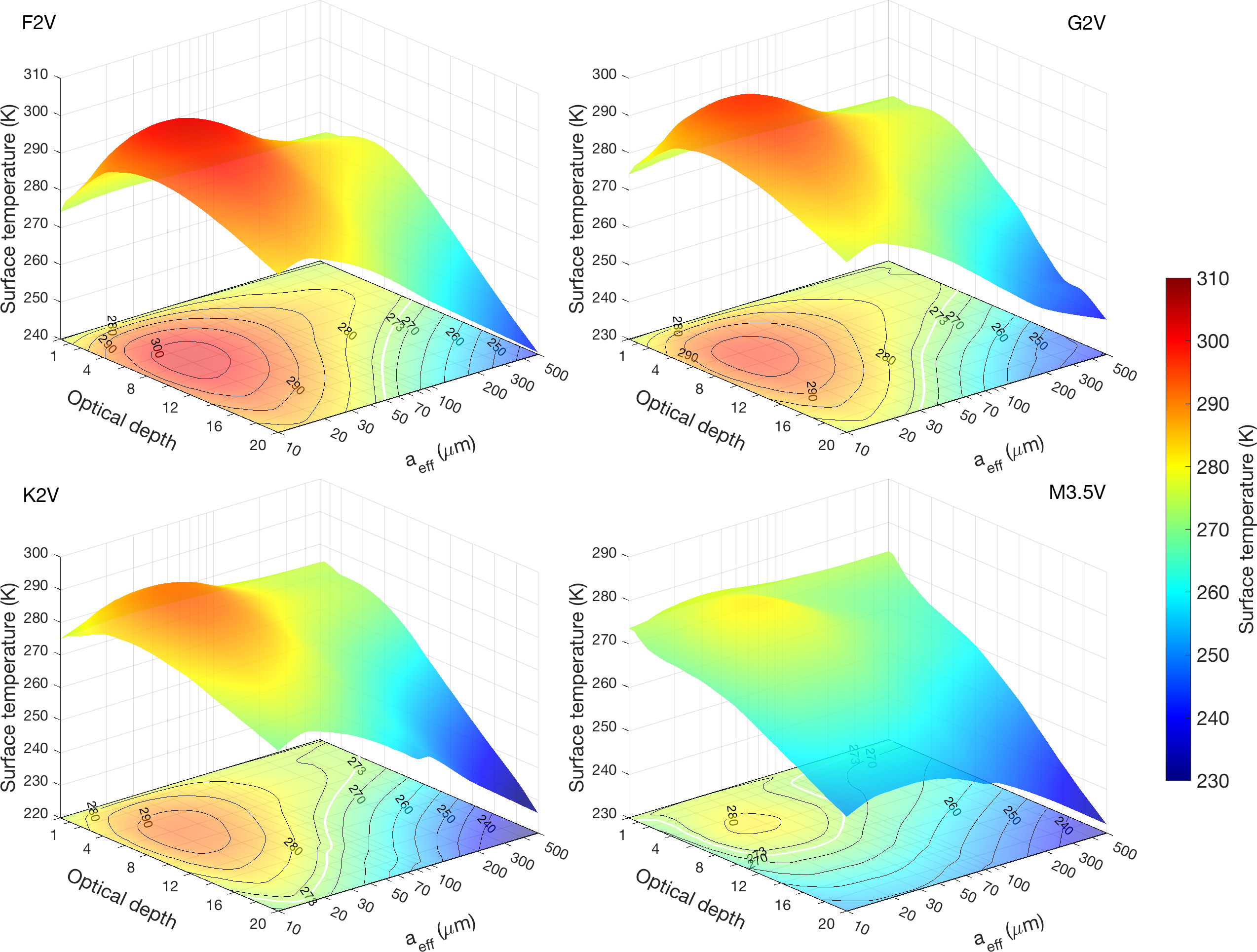}
  \caption{Impact of \ce{CO2} ice clouds on the surface temperature in atmospheres of planets around different main-sequence stars. Surface temperatures are shown as a function of 
optical depth (at a wavelength of $\lambda = 0.1 \ \mathrm{\mu m}$) and effective particle size of the \ce{CO2} ice clouds. The black contour 
lines are given in steps of 5 K. The white contour line indicates the clear-sky case.}
  \label{fig:surface_temp_2d_all}
\end{figure*}

\section{Climatic impact of \ce{CO2} ice clouds}
\label{sec:cloud_climatic_impact}

In this section, I first investigate the climatic effect of the \ce{CO2} clouds for planets orbiting different central stars. The aim is to test the impact of the stars' spectral energy 
distribution on the efficiency of the net greenhouse effect .
The results presented in \citet{Kitzmann2013A&A...557A...6K} suggest that the climatic impact is directly affected by the spectral distribution of the incident stellar radiation. 
M-stars, for example, seem to results in only a very small net greenhouse effect of the \ce{CO2} ice particles. This is caused by the shift of the stellar spectrum more towards the near- and mid-infrared compared to solar-type stars, such as our Sun. 
The shift of the spectrum to longer wavelengths makes Rayleigh scattering very inefficient and, thus, leads to a reduced net greenhouse effect by the cloud particles. 
However, \citet{Kitzmann2013A&A...557A...6K} were not able to quantify the impact on the surface temperature.

In order to be comparable, a common scenario is used for all central stars. In each case, the atmosphere is composed of six bar \ce{CO2} gas. Neither \ce{N2} nor \ce{H2O} are considered for these calculations.
The planets are placed at orbital distances, where a surface temperature of 273.15 K (freezing point of water) is obtained for each central star.
For these cases, the corresponding temperature-pressure profiles are shown in Fig. \ref{fig:clear_sky_profiles}, along with the saturation vapour pressure curve of \ce{CO2}.

In all cases, the troposphere is separated into two different convective regimes: A dry adiabatic region in the lower troposphere and a moist \ce{CO2} adiabatic temperature profile in the upper part.
This upper part indicates the atmospheric region, where a \ce{CO2} ice cloud could potentially form.

To study the impact of the \ce{CO2} clouds, a cloud layer is inserted into these saturated regions in the following. 
For the solar-type stars, the cloud layer is centred around 0.1 bar. 
This corresponds roughly to the position of the cloud layer in the studies of \citet{Forget1997} and \citet{Mischna2000Icar}.
In the case of the M-type dwarf, the cloud layer has to be located deeper in the atmosphere (0.4 bar) because the strong absorption of stellar radiation by the atmosphere leads to a warmer middle 
atmosphere and, thus, to a reduced tropopause height. The positions of the cloud layers is marked in Fig. \ref{fig:clear_sky_profiles}.

A supersaturated \ce{CO2}-dominated atmosphere would provide a large amount of condensible material. Thus, the \ce{CO2} ice particles could potentially reach large particle sizes.
The only detailed microphysical study by \citet{Colaprete2003JGRE} obtained particle sizes of up to 1000 $\mu$m for an early Martian atmosphere composed of two bar of \ce{CO2}. 
On the other hand, the more simple cloud schemes in the 3D GCM by \citet{Forget2013Icar..222...81F} resulted in particle sizes almost two orders of magnitudes smaller. 
One probable cause for this discrepancy could be the high, critical supersaturation (1.35, experimentally derived by \citet{Glandorf2002Icar}) used in \citet{Colaprete2003JGRE} to initiate the heterogeneous nucleation. Additionally, the clouds condense from the major atmospheric constituent meaning that a substantial amount of condensible material is available at this high supersaturation. This can result in the growth of very large ice particles.

To cover this parameter space, the effective particle sizes are varied between 10 $\mu$m and 500 $\mu$m in this study.
The resulting surface temperatures for optical depths of the cloud layer between 0.1 and 20 are shown in Fig. \ref{fig:surface_temp_2d_all} for all four central stars. The white contour lines indicate the clear-sky temperatures of 273.15 K.

The results presented in Fig. \ref{fig:surface_temp_2d_all} confirm the findings from \citet{Kitzmann2013A&A...557A...6K} and \citet{Kitzmann2016ApJ...817L..18K} with respect to the fact, that \ce{CO2} clouds are only effective in producing a net greenhouse effect within a small parameter range. The extent of this parameter range, on the other hand, depends crucially on the spectral type of central star; this is relatively large for F-type stars and very small for M-dwarfs.

On the other hand, the most effective particle radius for a net greenhouse effect is almost independent of the stellar type and is approximately 25 $\mu$m. As discussed in \citet{Kitzmann2016ApJ...817L..18K}, this is not too surprising because the scattering greenhouse effect occurs in the infrared wavelength region. In order to be effective scatterers in this wavelength region, the particle size must be comparable to the IR wavelengths.

The ability of the cloud layer to limit the loss of shortwave radiation to space due to molecular Rayleigh scattering, however, crucially depends on the incident stellar spectrum. The increased scattering of shortwave radiation between the lower 
atmosphere and the cloud base allows more radiation to be kept within the atmosphere than in the clear-sky case. Since the molecular Rayleigh scattering is proportional to $\propto \lambda^{-4}$, it is much less effective for cooler stars because the stellar spectrum is more shifted to near-infrared wavelengths. Thus, the stellar radiation of late-type stars is predominantly absorbed in the upper planetary atmosphere rather than being scattered. Consequently, the net heating effect of the \ce{CO2} ice clouds is considerably smaller in these cases.

\begin{figure}
  \includegraphics[scale=0.57]{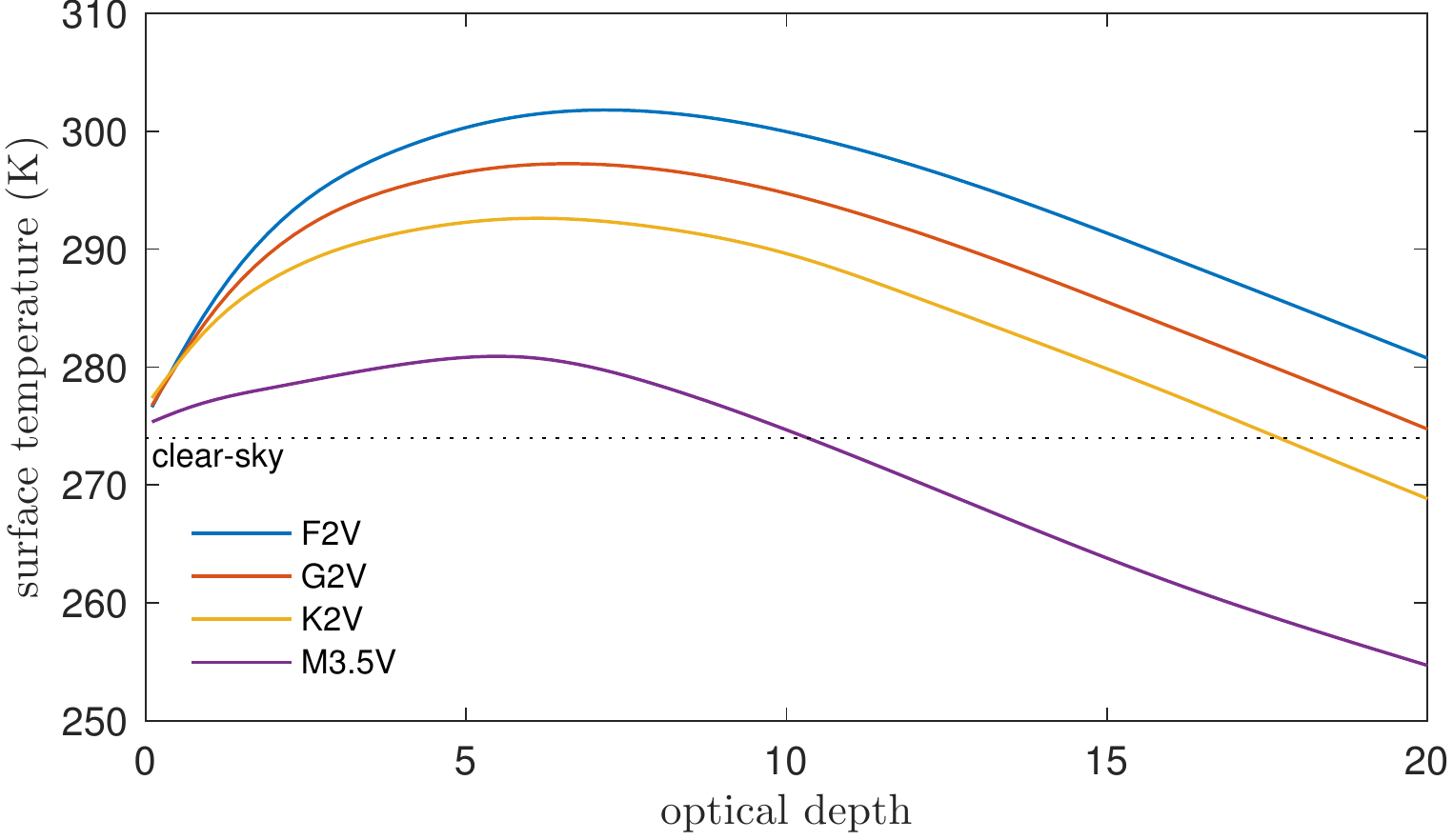}
  \caption{Surface temperatures as a function of the cloud's optical depth. 
           Results are shown for the four different central stars and an effective cloud particle radius of 25 $\mu$m.        
           The horizontal dotted line denotes the clear-sky case (273.15 K surface temperature).}
  \label{fig:surface_temp_25mu}
\end{figure}

Figure \ref{fig:surface_temp_25mu} shows the resulting surface temperature as a function of cloud layer's optical depth but for a fixed effective particle radius of 25 $\mu$m.
As already discussed in \citet{Kitzmann2016ApJ...817L..18K} for the conditions of the early Martian atmosphere, the \ce{CO2} clouds are only effective at optical depths of approximately 5 to 8. For larger optical depths, the net heating effect becomes increasingly smaller, up to the point where the presence of the cloud layer results in atmospheric cooling.
As already apparent from Fig. \ref{fig:surface_temp_2d_all}, the irradiance from an F-type star produces the largest net greenhouse effect, with a temperature of up to 30 K higher than the clear-sky 
case. The planet around the M-dwarf, on the other hand, doesn't benefit strongly from the cloud's greenhouse effect. Here, the largest temperature increases are at most 6 K.
On the contrary, for optical depths larger than 10, the \ce{CO2} ice cloud produces a net cooling effect that reaches temperature decreases of more than 20 K at an optical depth of 20.

These results will, of course, have a direct impact on the outer boundaries of the habitable zone around these different central stars. This is investigated further in the following section.

\section{Effect of \ce{CO2} clouds on the outer boundary of the habitable zone}
\label{sec:co2_hz}

\subsection{Cloud-free limit}

Before the impact of clouds is evaluated, I first perform calculations without the presence of clouds to obtain the cloud-free outer boundary of the habitable zone, as done by, for example, \citet{Kasting1993} or 
\citet{Kopparapu2013ApJ...765..131K}. These publications used an inverse modelling approach. This approach makes assumptions on the temperature profiles in the radiative equilibrium part of the 
atmospheres, which allows them to obtain the outer HZ boundary without iterating the model into thermal equilibrium.
Here, in contrast to that, I use full atmospheric calculations, that is, no a priori assumptions of the temperature profile are made and all results represent fully converged model calculations.

Additionally, \citet{Kasting1993} and \citet{Kopparapu2013ApJ...765..131K} used an increased surface albedo to account for the net cooling effect of water droplet and ice clouds. 
The surface albedo was tuned to yield a mean Earth surface temperature for an Earth-like planet around the Sun. 
Assuming that the impact of clouds neither depends on the central star nor on the atmospheric temperatures, this surface albedo was kept constant in all cases. 
\citet{Kitzmann2010A&A...511A..66K}, on the other hand, showed, that the greenhouse effect of the water ice clouds, for example, depends on the atmospheric temperatures.
For lower temperatures, this greenhouse effect has an increased efficiency, such that the tuned surface albedo would need to be decreased to capture this effect.

\begin{figure}
  \includegraphics[scale=0.57]{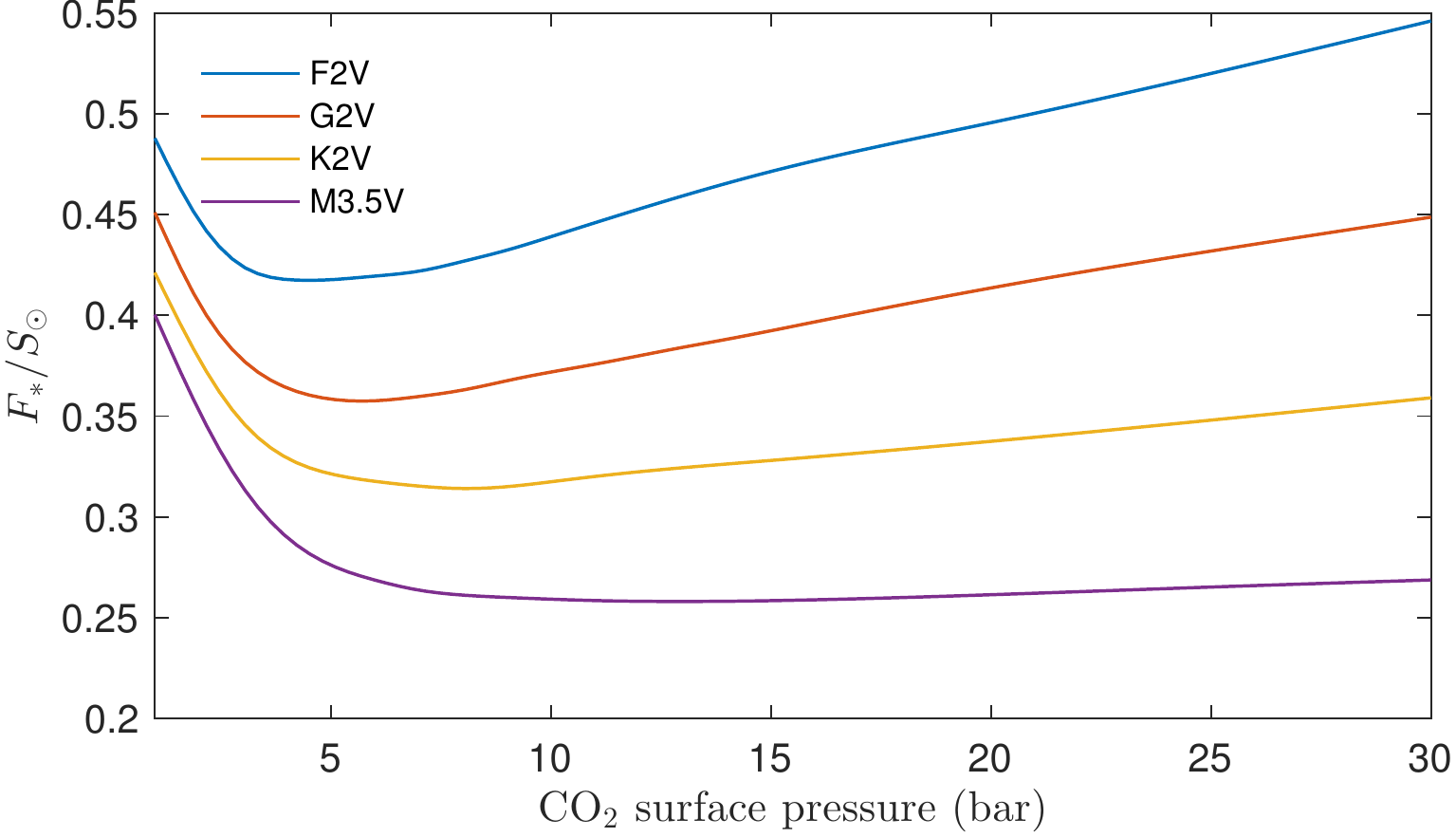}
  \caption{Maximum greenhouse effect of \ce{CO2} dominated atmospheres for planets orbiting different main-sequence stars. 
           The figure shows the stellar insolation required to obtain a temperature of 273.15 K as a function of the \ce{CO2} surface pressure. 
           The stellar flux at the top of the atmosphere $F_*$ is depicted in units of the solar constant $S_\odot$.
           The minima of the curves correspond to the maximum greenhouse effect and indicate the outer boundaries of the cloud-free habitable zone.}
  \label{fig:co2_var}
\end{figure}

However, since the outer boundary is determined for a thick \ce{CO2} atmosphere, the exact value of the surface albedo is not of great importance because most of the incident stellar radiation will 
be absorbed or scattered by the \ce{CO2} molecules before it reaches the surface. Only at low \ce{CO2} surface pressures should deviations be expected to occur.

Following the assumptions of \citet{Kasting1993} and \citet{Kopparapu2013ApJ...765..131K}, I use an atmosphere composed of \ce{N2}, \ce{CO2}, and \ce{H2O}. The amount of \ce{N2} is fixed at one bar, 
while the vertical distribution of \ce{H2O} is given by the relative humidity parametrisation of \citet{Manabe67}. The atmospheric \ce{CO2} content, on the other hand, is a free parameter.
By varying the amount of stellar insolation, I obtain the orbital distances, where the assumed planet possesses a surface temperature of 273.15 K for a given surface pressure of \ce{CO2}. The corresponding results for all four central stars are presented in Fig. \ref{fig:co2_var}.

The results clearly show the well-known maximum greenhouse effect of \ce{CO2} \citep{Kasting1993}. Carbon dioxide is only an efficient greenhouse gas below a certain maximum partial pressure. 
For surface pressures higher than this value, the greenhouse effect is offset by the molecular Rayleigh scattering, such that the net effect of the \ce{CO2} molecules would be a dominating cooling effect.
This maximum greenhouse effect is a function of the spectral distribution of the incident stellar radiation. It is more dominant for an F-type star and less effective for late-type stars because their 
spectra are more shifted towards the near infrared, which makes Rayleigh scattering rather inefficient. The results of Fig. \ref{fig:co2_var} agree overall with the values of 
\citet{Kopparapu2013ApJ...765..131K}.
Most deviations can be found at lower \ce{CO2} surface pressures, where the different value of the surface albedo influences the results. 
Figure \ref{fig:co2_var} allows us to obtain the cloud-free outer boundary of the habitable zone, through use of the minima of the curves, that is, the smallest possible insolations.

\begin{figure}
  \includegraphics[scale=0.57]{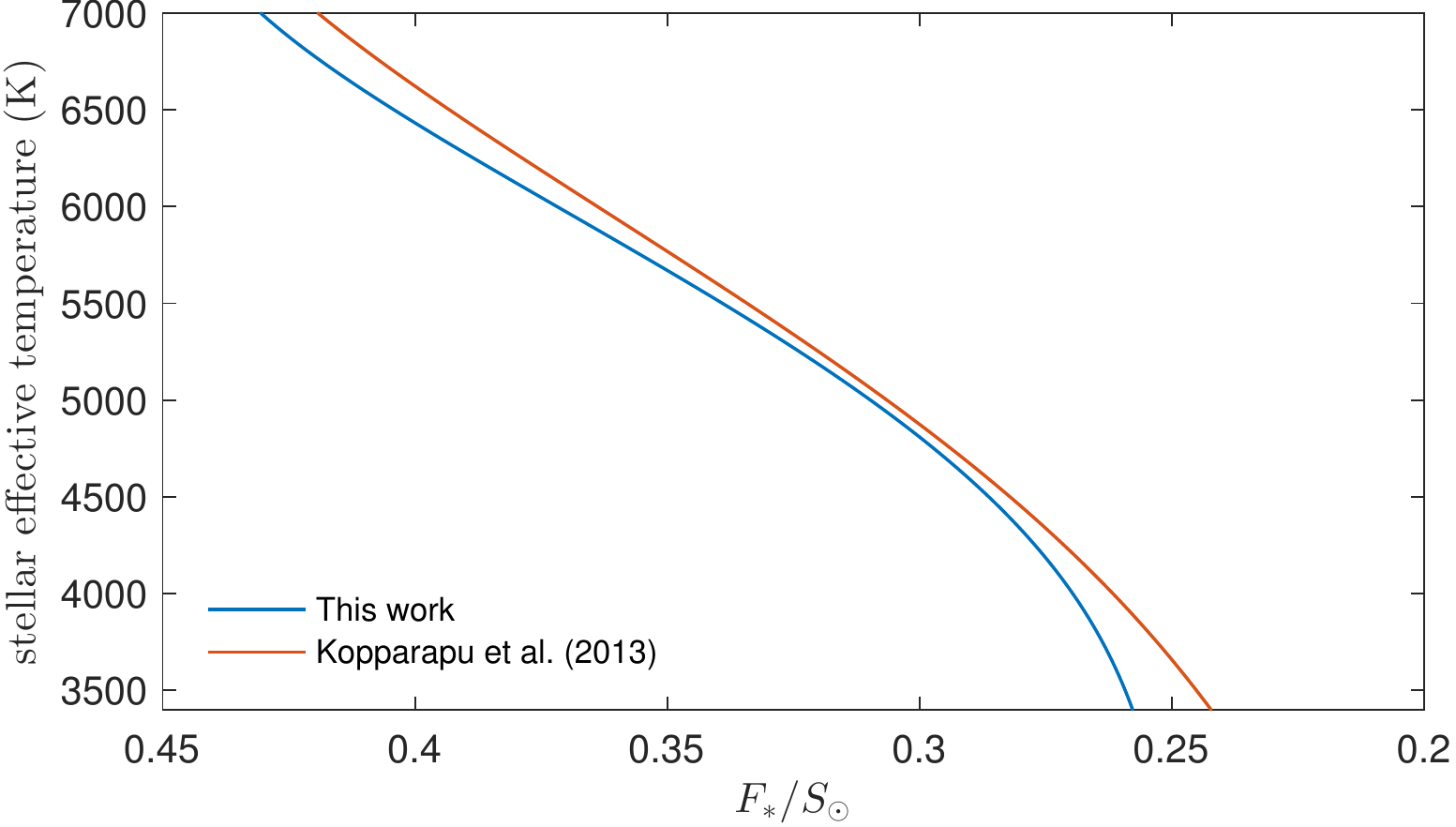}
  \includegraphics[scale=0.57]{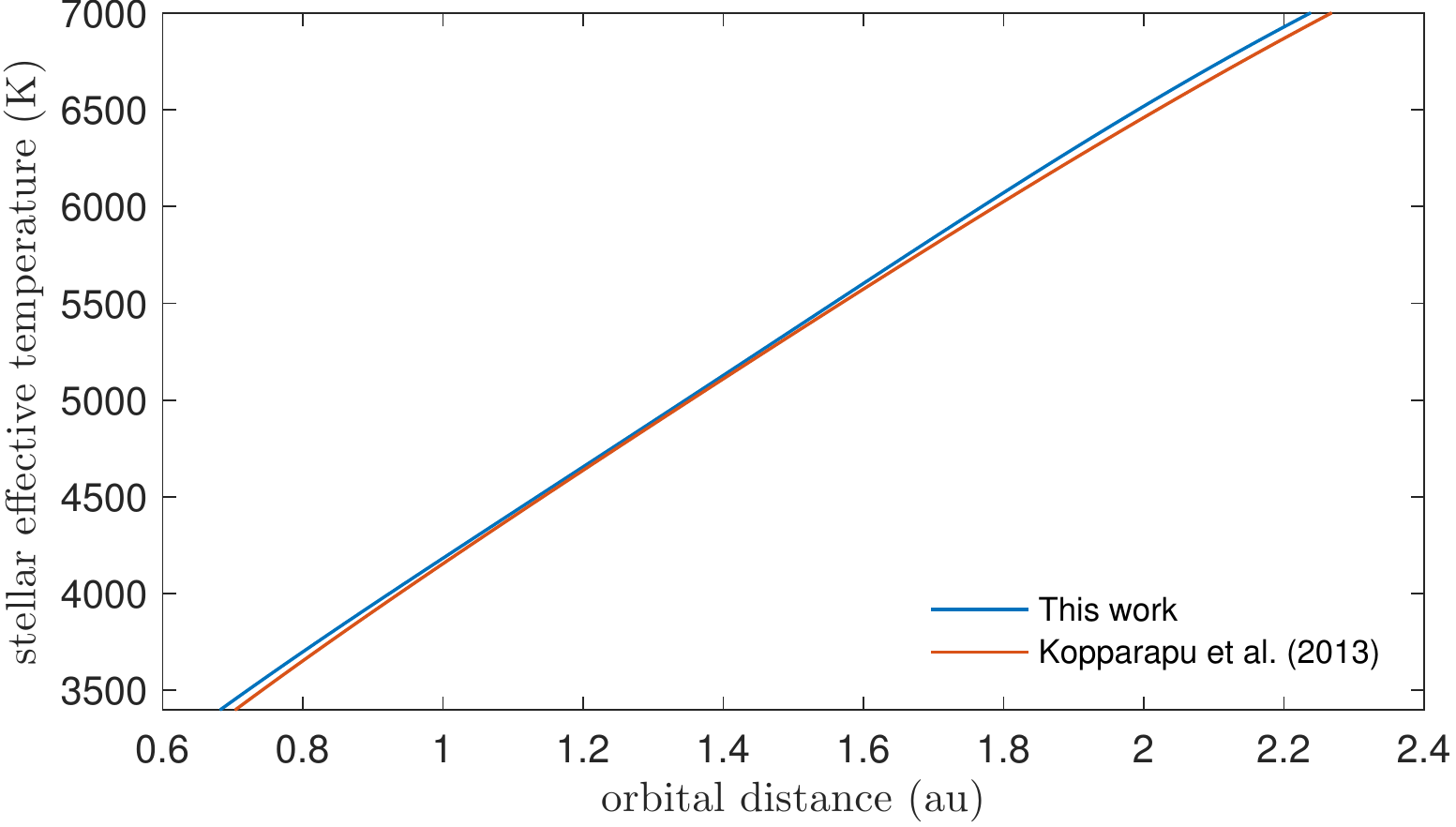}
  \caption{Cloud-free outer boundary of the habitable zone. The critical insolations $F_*$ in units of the present day solar constant $S_\odot$ for the maximum greenhouse effect (\textit{upper 
panel}) and the corresponding orbital distances (\textit{lower panel}) of the outer HZ boundary are shown as a function of the stellar effective temperature. The results of 
\citet{Kopparapu2013ApJ...765..131K} are additionally presented for comparison (red curves).}
  \label{fig:hz_seff}
\end{figure}

These critical insolations for the maximum greenhouse effect are shown in  Fig. \ref{fig:hz_seff} (upper panel) as a function of the stellar effective temperature. 
Additionally, the figure depicts the corresponding results from \citet{Kopparapu2013ApJ...765..131K}. It should be noted that \citet{Kopparapu2013ApJ...765..131K} use a value of 1360 W m$^{-2}$ for 
$S_\odot$, whereas here a solar constant of 1366 W m$^{-2}$ is employed.

The deviations found in terms of the stellar insolation at the outer boundary as a function of $T_\mathrm{eff}$ are rather small. 
These minor differences can be explained by the use of different surface albedos and the employed radiative transfer schemes.

The outer boundary position can also be expressed in terms of orbital distance $d$ by the usual relation $d = \sqrt{S_\odot / F_*}$, where $F_*$ is the stellar insolation at the top of the atmosphere and $\left[ d \right] = \mathrm{au}$. 
The resulting cloud-free distances are shown in the lower panel of Fig. \ref{fig:hz_seff}, again with the outer HZ boundary determined by 
\citet{Kopparapu2013ApJ...765..131K} for comparison. 
Figure \ref{fig:hz_seff} suggests, that the differences in terms of orbital distances between the two different studies are relatively small and almost negligible, considering the different modelling approaches. As noted in the introduction, the revised orbital distances by \citet{Kopparapu2013ApJ...765..131K} also differ slightly from the original values obtained by \citet{Kasting1993}.

Following the approach of \citet{Selsis2007}, I perform a polynomial fit of the resulting orbital distances as a function of the stellar effective temperature.
The results for the outer HZ boundary locations in astronomical units are expressed by a 4th-order polynomial of the form:
\begin{equation}
  d = c_1 T_\mathrm{eff} + c_2 T_\mathrm{eff}^2 + c_3 T_\mathrm{eff}^3 + c_4 T_\mathrm{eff}^4 \ , \quad \text{with} \ [d] = \mathrm{au} \ . 
  \label{eq:hz_fit}
\end{equation}
The corresponding parameters $c_i$ are given in Table \ref{table:hz_fit}.

\begin{table}[h]
    \caption[]{Coefficients for the outer HZ boundary in the clear-sky and 100\% cloud limit to be used in Eq. \ref{eq:hz_fit}.}
    \label{table:hz_fit}
    \centering
    \begin{tabular}{l r r}   
        \hline
        \noalign{\smallskip}
          & Clear-sky &  100\% clouds\\
        \hline
        \noalign{\smallskip}
        c$_1$ (au/K) & -2.207e-04 & -2.453e-04 \\
        c$_2$ (au/K$^2$) & 2.063e-07 & 2.075e-07 \\
        c$_3$ (au/K$^3$) & -2.968e-11 & -2.503e-11 \\
        c$_4$ (au/K$^4$) & 1.603e-15 & 1.181e-15\\
        \hline
        \noalign{\smallskip}
    \end{tabular}
\end{table}

\subsection{Impact of \ce{CO2} clouds}

In this subsection the impact of the \ce{CO2} ice clouds on the position of the outer HZ boundary is calculated.
The calculations are restricted to the most optimum cases, that is, cloud optical depth and particle radii as well as the \ce{CO2} surface pressure are chosen such that the highest possible net heating effect of the \ce{CO2} ice clouds is obtained.
Additionally, in accordance with Sect. \ref{sec:cloud_climatic_impact}, the cloud coverage is set to 100\%.
Thus, the values for the outer boundary presented in this section represent the upper limits of the impact of \ce{CO2} ice cloud on the outer HZ location.

The effective radius of particle size distribution is chosen based on the results from the previous subsection.
The results suggest, that the most effective particle size is independent from the central type and given by approximately $a_\mathrm{eff} \approx 25 \, \mu\mathrm{m}$.
This particle size will therefore be used in the following for all calculations.

For the most optimum cases, the atmospheric \ce{CO2} partial pressure and the optical depths of the clouds are slightly higher than what has been presented in the previous section. For example, the maximum greenhouse effect of the clear-sky atmosphere in the G-star case occurs around a \ce{CO2} partial pressure of approximately 5.7 bar (see Fig. \ref{fig:co2_var}), while the most effective optical 
depth of the cloud layer in the 6 bar pure \ce{CO2} atmosphere is approximately 6.5.
When varying both, the \ce{CO2} content and the cloud optical thickness, the highest impact on the surface temperature occurs for a carbon dioxide surface pressure of 8 bar and an optical depth of 8. 
However, the difference between these two cases in terms of the stellar insolation required to obtain a surface temperature of 273.15 K is only about 0.008 $S_\odot$ and, thus, has no impact on the position of the outer HZ boundary.

\begin{figure}
  \includegraphics[scale=0.57]{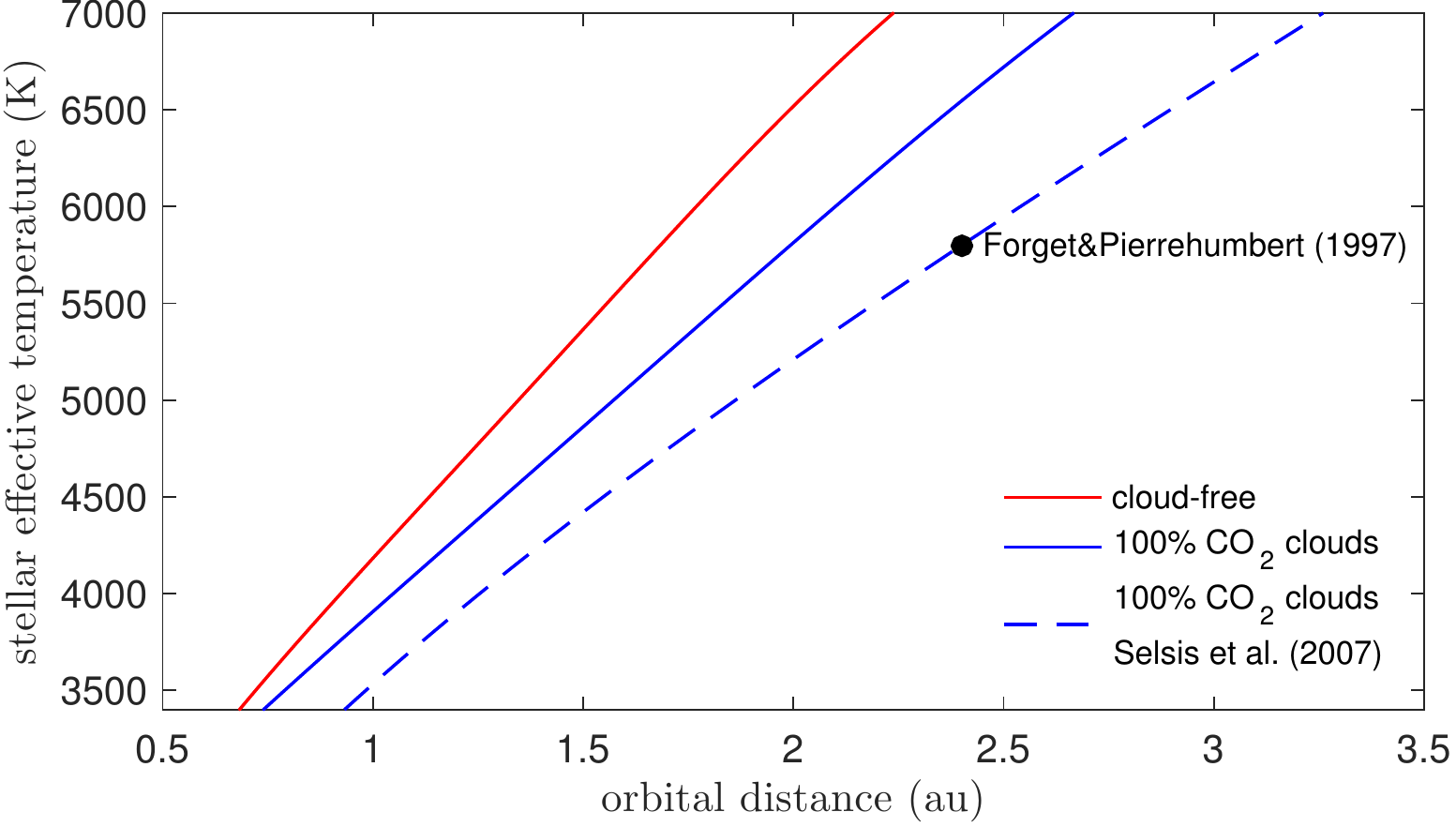}
  \caption{Impact of \ce{CO2} ice clouds on the position of the outer HZ boundary. The solid red curve shows the cloud-free boundary, while the solid blue curve shows the effect of clouds on the habitable zone. 
The results from \citet{Selsis2007} (dashed, blue curve) and \citet{Forget1997} for the impact of clouds on the HZ boundary are additionally shown for comparison}
  \label{fig:hz_clouds}
\end{figure}

The orbital distances of the HZ boundary for all four central stars are again fitted with the polynomial in Eq. \eqref{eq:hz_fit}, with the resulting parameters given in Table \ref{table:hz_fit}.
The results of this polynomial fit are shown in Fig. \ref{fig:hz_clouds} along with the results from \citet{Selsis2007} and \citet{Forget1997} for comparison.

The increase of the orbital distances compared to the cloud-free case is clearly a function of the stellar effective temperature. For a planet around cool M-dwarfs, the increase in distance is relatively small. In the best-case scenario presented in this section, the increase is only approximately 0.05 au. Given the fact that, in reality, the atmospheric and \ce{CO2} cloud properties will be less than 
optimal, one shouldn't expect a strong positive effect of carbon dioxide clouds in these cases. On the contrary, for less than optimal cloud properties, the outer boundary could even be shifted to smaller orbital distances than the clear-sky case for cooler central stars. 
For stars with higher effective temperatures, the increase in orbital distance can be up to 0.5 au. For the F-type star, for example the presence of \ce{CO2} ice clouds would allow the planet to be located at 2.54 au, which is 0.44 au farther than the clear-sky value of 2.1 au.

As mentioned in the introduction, the only atmospheric modelling calculations for the impact of the carbon dioxide clouds on the position of the outer HZ boundary was done by \citet{Forget1997}, 
who state a distance of 2.4 au for the Sun. This result has been used by \citet{Selsis2007} to scale the cloud-free distances from \citet{Kasting1993}. It should be noted though, that the absorption coefficients for \ce{CO2} molecules in \citet{Kasting1993} and \citet{Forget1997} were overestimated because the continuum absorption was accounted for twice in their atmospheric model \citep[see][for details]{Wordsworth2010Icar..210..992W}.
The less effective greenhouse effect by the \ce{CO2} ice clouds in this study results in a much smaller value for the orbital distance in this case. According to Fig. \ref{fig:hz_clouds}, the 
orbital distance for a G2V star is about 2 au and, thus, 0.4 au smaller than what has previously been estimated. 
Compared to the results from \citet{Selsis2007}, the new estimates provided by this work shift the boundary towards smaller orbital distances by approximately 0.5 au for F-type stars and 
approximately 0.2 au for M-dwarfs.

\section{Summary}
\label{sec:summary}

In this study the climatic effects of CO$_2$ ice clouds in CO$_2$-dominated atmospheres of terrestrial planets around different types of main-sequence dwarf stars and the impact of \ce{CO2} clouds on the location of the outer boundary of the classical habitable zone are investigated. 
A radiative-convective atmospheric model employing an accurate discrete ordinate radiative transfer was used to calculate surface temperatures for a broad range of particle sizes and optical depths of CO$_2$ ice particles. 

As a first step, the climatic impact of \ce{CO2} clouds in a thick, \ce{CO2}-dominated atmosphere was evaluated for different stellar spectral energy distributions and as a function of the effective particle radius and the cloud's optical depth.
As already pointed out by \citet{Kitzmann2013A&A...557A...6K}, the heating and cooling effect should be a direct function of the incident stellar spectrum. 
The results from this work suggest that for these thick, \ce{CO2}-dominated atmospheres the cloud's radiative forcing yields, at most, temperature increases smaller than 6 K for cool M-stars. On the other hand, central stars with higher effective temperatures resulted in surfaces temperatures approximately 30 K higher than the corresponding clear-sky case.
Furthermore, the parameter range where a \ce{CO2} ice cloud can have a net positive effect is rather limited for late-type stars. A heating effect is only obtained in a small parameter range with respect to optical depths and effective particle sizes. Outside this parameter range, the clouds would have a dominating albedo effect, thus cooling the surface.
The most effective particle size for the scattering greenhouse effect was found to be approximately 25 $\mu\mathrm{m}$, independent from the central star type.

Following this, the impact of dry ice clouds on the position of the classical HZ's outer boundary was investigated. 
Atmospheric calculations without clouds were performed first to obtain the cloud-free habitable zone boundary. The results of this cloud-free limit are very similar to the orbital distances published by \citet{Kopparapu2013ApJ...765..131K}. 

The outer boundary influenced by clouds has been calculated for the most optimum scenarios and, thus, represents an upper limit. The additional greenhouse effect allows the orbital distance of the outer HZ boundary to be increased by up to 0.5 au for F-stars. Stars with lower effective temperatures, on the other hand, don't benefit strongly from the \ce{CO2} ice clouds. The less effective net greenhouse effect results in an extension of only 0.05 au.

Compared to the orbital distance of the outer HZ boundary of 2.4 au for the Sun published by \citet{Forget1997}, which has been obtained using a simplified two-stream method, the revised distance obtained by the more accurate radiative transfer treatment in this study is 0.4 au smaller. Thus, all parametrisations of the outer boundary which rely on that value (e.g. \citet{Selsis2007} or \citet{Kaltenegger2011ApJ}) should be revisited.

It should be noted that the outer HZ boundary calculated in this work only considers the classical \ce{CO2}-dominated atmospheres. The presence of other greenhouse gases, such as \ce{CH4}, might also lead to an additional extension of the HZ. Other possible mechanisms also include, for example, the greenhouse effect provided by a \ce{H2}-dominated atmosphere as studied by e.g. \citet{Pierrehumbert2011ApJ}, for example.    

There are also still open questions remaining regarding the climatic impact of \ce{CO2} clouds. Models including a treatment for \ce{CO2} ice cloud formation have so far strongly disagreed on the resulting particle sizes \citep[e.g.][]{Forget2013Icar..222...81F, Colaprete2003JGRE}, such that additional research in this area is warranted. Additionally, all models have, so far, assumed spherical particles and used Mie theory to obtain the optical properties of the cloud particles. 
However, as shown by laboratory measurements of \citet{Behnken1912}, \citet{Wahl1913}, or \citet{Wergin1997} CO$_2$ ice crystals can have cubic or octahedral shapes. 
Combinations of both (cuboctahedra) or more complicated shapes such as rhobic-dodecahedral crystals can also occur.
Since the greenhouse effect of carbon dioxide clouds is determined by the scattering properties of the ice particles, their shapes should have a large impact on the resulting optical properties (e.g. scattering phase function) and, thus, also on the climatic impact.

\begin{acknowledgements}
      D.K. gratefully acknowledges the support of the Center for Space and Habitability of the University of Bern and the MERAC Foundation for partial financial assistance.
      This work has been carried out within the frame of the National Centre for Competence in Research PlanetS supported by the Swiss National Science Foundation. D.K. acknowledges the financial support of
      the SNSF.
\end{acknowledgements}

\bibliographystyle{aa}
\bibliography{references}

\end{document}